\documentclass[letterpaper, 10 pt, conference]{ieeeconf}  

\usepackage{comment}
\usepackage{cite}
\usepackage{graphicx}
\usepackage{booktabs} 

\IEEEoverridecommandlockouts                              

\title{\LARGE \bf
PCNNA: A Photonic Convolutional Neural Network Accelerator
}

\author{ \parbox{7 in}{\centering Armin Mehrabian, Yousra Al-Kabani, Volker J Sorger, and Tarek El-Ghazawi
         \\
         {\small Department of Electrical and Computer Engineering\\
         The George Washington University, Washington, DC, USA}\\
         {\small \{armin,yousra,sorger,tarek\}@gwu.edu}}
         \hspace*{ 0.5 in}
         \thanks{\copyright 2018 IEEE}\\
}

\begin{document}

\maketitle
\pagestyle{empty}
\begin{abstract}
Convolutional Neural Networks (CNN) have been the centerpiece of many applications including but not limited to computer vision, speech processing, and Natural Language Processing (NLP). However, the computationally expensive convolution operations impose many challenges to the performance and scalability of CNNs. In parallel, photonic systems, which are traditionally employed for data communication, have enjoyed recent popularity for data processing due to their high bandwidth, low power consumption, and reconfigurability. Here we propose a Photonic Convolutional Neural Network Accelerator (PCNNA) as a proof of concept design to speedup the convolution operation for CNNs. Our design is based on the recently introduced silicon photonic microring weight banks, which use broadcast-and-weight protocol to perform Multiply And Accumulate (MAC) operation and move data through layers of a neural network. Here, we aim to exploit the synergy between the inherent parallelism of photonics in the form of Wavelength Division Multiplexing (WDM) and sparsity of connections between input feature maps and kernels in CNNs. While our full system design offers up to more than 3 orders of magnitude speedup in execution time, its optical core potentially offer more than 5 order of magnitude speedup compared to state-of-the-art electronic counterparts.
\end{abstract}
\section{INTRODUCTION}

CNNs have been able to reach record-breaking performance in many tasks including but not limited to computer vision \cite{he2016deep}, speech recognition \cite{hannun2014deep}, and NLP \cite{chen2017reading}. However, the success comes at the cost of high computational demands. Convolution operations account for roughly 90\% of the total operations in a CNN \cite{cong2014minimizing}. While CNNs enjoy highly-parallel operations within a layer, data dependencies across layers challenge any attempt of inter-layer parallelization. The latter also poses scalability burdens in terms of power and throughput. Such challenges are even more magnified when electronics seem to be hitting fundamental power and speed limitations.

Photonics is considered as a promising alternative to electronics both for communication and more recently for computation \cite{miller2017attojoule}. Photonic systems offer inherent parallelism potentials through their Wavelength Division Multiplexing (WDM) capability. Low Light Matter Interaction (LMI) makes photonics a great choice for signal transmission to far distances with minute loss and energy. In addition, the linear nature of light can be exploited to perform linear mathematical operations such as multiplication and addition. This makes photonics an appealing choice for the implementation of CNNs due to their heavy reliance on MAC operation.

Previous work on CNN inference hardware has primarily focused on electronic implementations, either in the form of FPGA \cite{zhang2015optimizing,suda2016throughput} or ASIC designs \cite{han2016eie,shafiee2016isaac}. Despite some recent efforts to implement neural networks using optics \cite{tait2017neuromorphic,shen2017deep}, to authors knowledge optical realization of CNNs has not been explored yet.

In this paper we propose a photonic convolution accelerator for CNNs inference-mode based on the recently proposed Micro-Ring-Resonator(MRR) banks and the broadcast-and-weight protocol \cite{tait2017neuromorphic}. We summarize the main contributions of this work as:
\begin{itemize}
\item A first proposed design for an optical CNN based on the recently proposed broadcast-and-weight protocol with MRR weight banks.
\item An optimization technique is proposed to reduce the number of microrings for the optical CNN realization.
\item An analytical framework is introduced to identify the number of microrings for any CNN layer and to estimate the execution time.
\end{itemize}

\section{Convolutional Neural Networks (CNNs)} CNNs' architecture enable them to receive inputs of higher dimensional shape and construct a hierarchy of feature representations. High-dimension inputs are inspected for presence of features (kernels) learned during a training process. This inspection process is carried out through a series of 4D convolution operations.
Thus, the output of convolution layer is a condensed set of values indicating the presence or absence of features in the input. For that, the outputs of layers in CNN is referred to as feature map. Unlike dense fully-connected layers a kernel in CNN only observes and operates on a narrow window of inputs referred to as receptive field. The size of this window equals the size of the kernel. Moreover, the connections between the kernel values and the input feature map receptive field is one-to-one rather than all-to-all. In other words, in CNNs convolutional layers enjoy sparsity of connections between input feature maps and kernels. This sparsity results in interesting characteristics including input feature map reuse. As the name suggests, a single feature map is reused as the convolution input for many different kernels. In this paper we will exploit this property of CNNs as the conceptual foundation of our design.

\section{Photonic Microring Resonator (MRR) banks}
\begin{figure}[ht]
    \centering
    \includegraphics[scale=0.22]{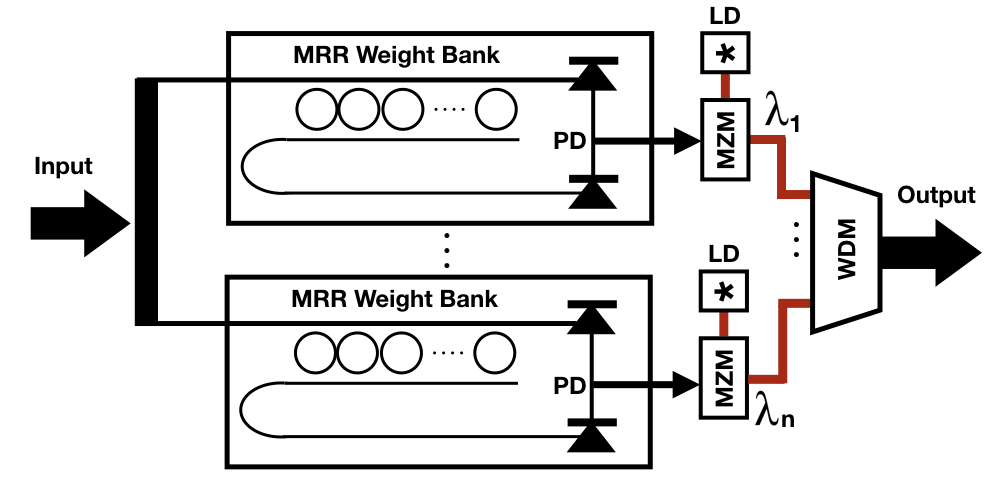}
    \caption{Broadcast-and-weight protocol using MRR banks. Incoming bundled wavelengths propagate through the MRR banks. Each bank weights each wavelength by changing the tune of corresponding rings. A photodiode sums up all the wavelengths into a photo-current. The photo-current modulates a laser beam of wavelength $\lambda_m$. All outgoing laser beams are multiplexed together to be broadcast to the next layer.}
    \label{fig:MRR bank}
\end{figure}

In \cite{tait2017neuromorphic} authors proposed a photonic ANN design based on the broadcast-and-weight protocol. Figure \ref{fig:MRR bank} shows the conceptual design of the broadcast-and-weight protocol. In this model MRR banks and photodiodes perform MAC operations while broadcast-and-weight protocol carry MAC results across layers. In broadcast-and-weight protocol each neuron output is multiplexed onto a distinct light wavelength using Laser Diodes (LD). Multiplexed wavelengths are bundled together and placed on a waveguide to broadcast to the destination layer. At the destination layer, each neuron receives all the incoming wavelengths. Each wavelength is then multiplied in amplitude with its corresponding microring. Multiplication is carried out by tuning rings in and out of resonance to a respective laser wavelength. Later, a photodiode sums up all the incoming wavelengths into an aggregate photo-current.

\section{Design Methodology} \label{sec:design_method}
PCNNA takes advantage of MRR banks proposed in \cite{tait2017neuromorphic} to perform MAC operations. However, one major drawback of this scheme is that the  number of microrings required to perform the multiplication part of MAC in each layer scales with $N_{i}\times N_{i+1}$, where $N$ is the number of nodes in a layer and $i$ is the layer number. This exponential increase in the number of microrings makes implementation of such networks challenging.
\begin{table}
\centering
  \caption{Summary of convolution layer parameters used in this work}
  \label{tab:conv_parameters}
  \begin{tabular}{cc}
    \toprule
    Parameter&Description\\
    \midrule
    $n$& Input feature map height and width\\
    $m$& Kernel height and width\\
    $p$& Padding size\\
    $s$& Stride step size\\
    $n_{c}$& Input feature map number of channels\\
    $N_{input}$& Input feature map size\\
    $N_{ouput}$& Output feature map size\\
    $N_{kernel}$& Size of kernel\\
    
\end{tabular}
\end{table}
\begin{figure*}[t]
    \centering
    \includegraphics[scale=0.46]{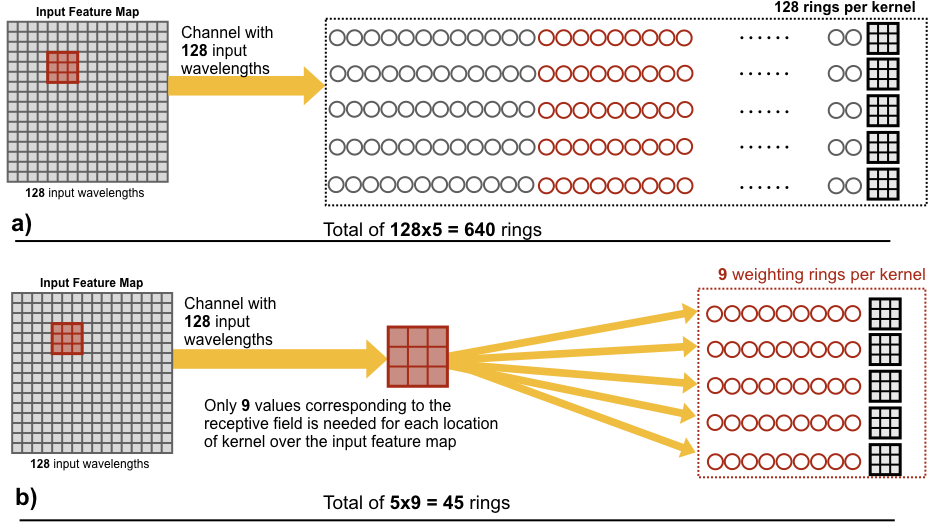}
    
\caption{MRR bank for an input feature map of size $16 \times 16$ and 5 kernels of size $3 \times 3$, a) without filtering the input feature map and b) with input feature map filtered to only pass through receptive field. It can be seen that taking advantage of narrow receptive field results in less number of total microrings.
}
    \label{fig:rings_concept}
\end{figure*}
For a kernel of size $N_{kernel}$, at each location of the kernel over the input feature map, only $N_{kernel}$ values corresponding to the receptive field of input feature map take part in the convolution. Therefore, using MRR banks, we only need to allocate $N_{kernel}$ microrings for weighting and the rest of $N_{input}-N_{kernel}$ values can be ignored. We will refer to these values as non-receptive field values for the rest of this paper. This ignoring of the non-receptive field values result in large savings in terms of both number of wavelengths that represent the input feature map and the number of microrings required in the following layer to demultiplex incoming wavelengths. Figure \ref{fig:rings_concept} conceptually shows the effect of filtering non-receptive field values in MRRs for convolution operation (a) with no filtering of the non-receptive field values and (b) with non-receptive field values filtered.

Current CNNs comprise of tens, if not hundreds, of layers with almost the same range of kernels per layer \cite{krizhevsky2012imagenet,szegedy2015going,he2016deep}. While filtering non-receptive field weight result in large savings in the number of microrings and respectively power consumption, implementing a full CNN using MRRs still requires large footprints and power consumptions. Therefore, in PCNNA we construct our design based on implementation of a single layer of CNN and virtually reusing it sequentially for different layers.

\begin{figure}[ht]
    \centering
    \includegraphics[scale=0.21]{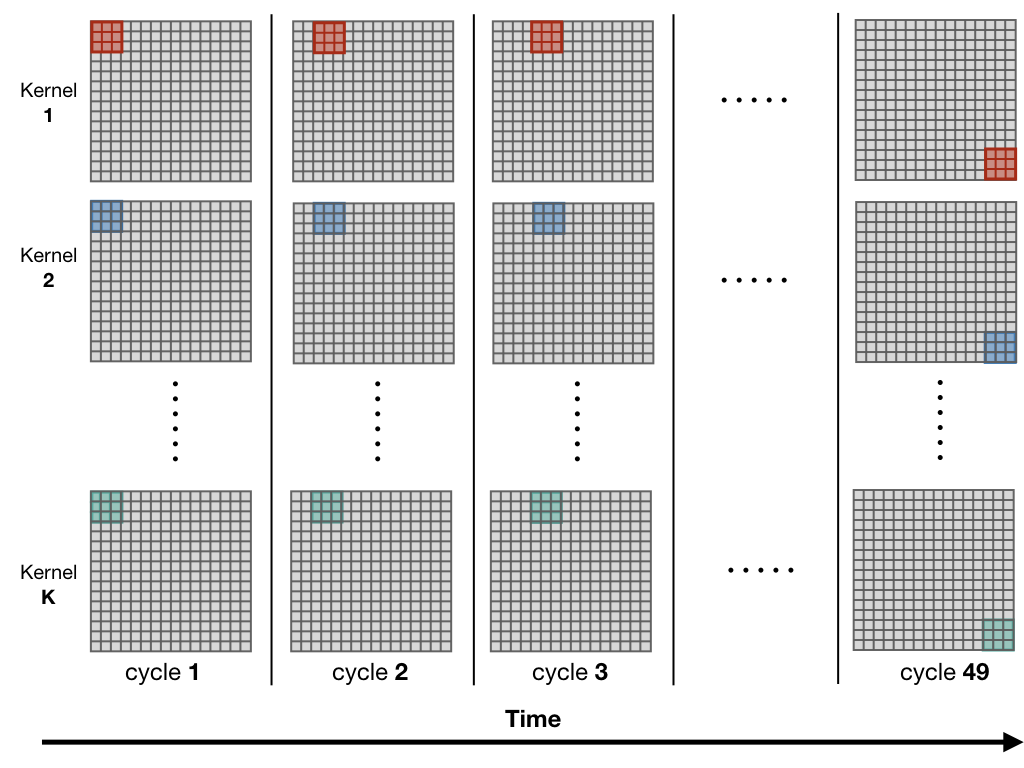}
    \caption{Parallel execution of kernel as kernels sequentially progress over various locations of the input feature map.}
    \label{fig:kernel_progress}
\end{figure}
In PCNNA convolution layers are processed sequentially. Convolution result values of each layer are stored back to the off-chip DRAM. In addition, for each location of the kernels corresponding to a receptive field, partial convolutions are processed sequentially at the optical core of PCNNA. But, as multiple kernels share the same receptive field values, convolution computations for different kernels are performed in parallel. Figure \ref{fig:kernel_progress} illustrates the parallel execution of \textbf{$K$} kernels as they progress across the input feature map.

\begin{figure}[ht]
    \centering
    \includegraphics[scale=0.23]{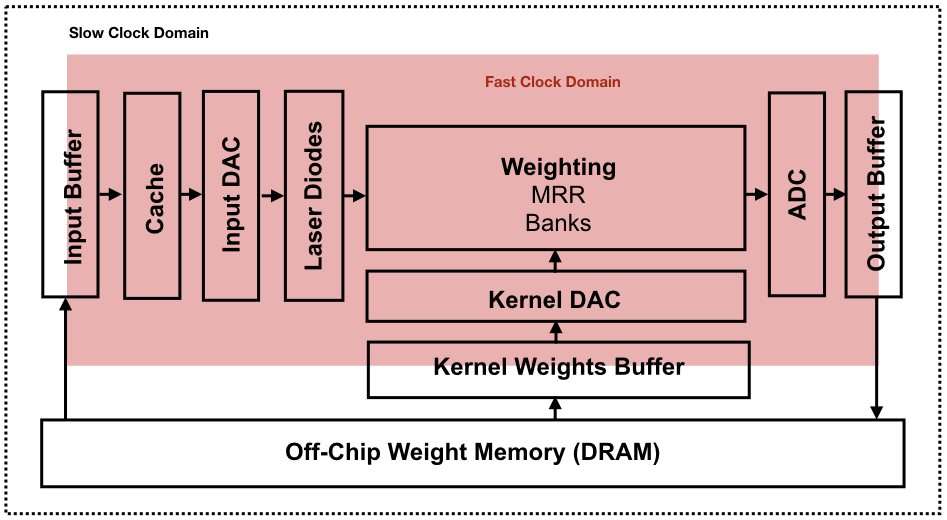}
    \caption{Conceptual high-level hardware architecture PCNNA. The shaded areas, which include the core optical components run of a fast clock (5GHz). Buffers isolate the fast optical core from the outside slow clock environment.}
    \label{fig:overall_architecture}
\end{figure}

At the high level PCNNA runs on two clock domains, a fast clock domain (5GHz), which runs the optical sub-systems and their immediate electronic circuitry, and a main slower clock domain to interface with the external environment. Figure \ref{fig:overall_architecture} depicts the overall hardware architecture of our design. PCNNA consist of a weighting MRR bank repository, which its microrings tune to kernel weights. Kernel weights are initially stored in an off-chip DRAM memory. Upon arrival of a new layer request, kernel weights are loaded from the off-chip DRAM into the Kernel Weights Buffer. Digital buffered weights are then converted into analog voltages, which control the tuning of the microrings in the MRR banks. 

Similarly, the input feature maps are initially stored in the off-chip DRAM. It should be noted that over the execution of one layer of a CNN the kernel weights do not change. However, due to sequential progression of kernels over the input feature maps, the input values are regularly updated. For instance, in Figure \ref{fig:kernel_progress} the values corresponding to kernels do not change until a new layer is loaded, but the input receptive field goes through 49 cycles. 

For each receptive field corresponding to a particular location of kernel a subset of input feature map values ($N_{kernel}$) are loaded into the Input Buffer. These buffered values are then moved closer to the PCNNA core and stored in small but fast cache memory. Each value in the cache memory is then converted to an analog signal using the input Digital to Analog Converters DACs. Laser beams of different wavelengths generated by Laser Diodes (LD) modulate the analog signal from the DAC. Laser beams fly through the MRR banks and their output photodiodes, which perform the multiplication and summation operations respectively. This process can be done quite fast (flight time of light) once all input values are loaded and converted to analog signals. Even the integrated photodiodes' operating frequency at 0 bias voltage can be as high as tens of GHz if not hundreds \cite{fossum2014review}. Hence, the whole optical weighting and summation fits within a single clock cycle of our fast clock domain. Lastly, computed convolution values are digitized back through the ADC and stored into the off-chip DRAM.

The latter procedure repeats for every location of kernels across the input feature map. For any consecutive location of kernels within a layer, only a fraction of input feature map values proportional to the size of the stride is required to be loaded into the cache. As a result, the bandwidth required for continuous loading of various receptive field values of input feature maps is minimized.

\section{Evaluations}
In this section we develop analytical models for the evaluation of our design in term of the area consumed by microrings and the execution time.

\subsection{Microring Area} For simplicity we assume that  input feature maps are square-face volumes. As a result, the size of input feature maps and kernels are as follows,
\begin{equation}
N_{input}=n\times n\times n_c
\end{equation}
\begin{equation}
N_{kernel}=m\times m\times n_c
\end{equation}
where $n$ and $m$ are the size of input feature map and kernel in x and y direction, and $n_c$ is the number of channels. For a given input feature map and a set of \textbf{$K$} kernels, the output feature map will have the size,
\begin{equation}
N_{output}=(\left[\frac{(n+2p-m)}{s}\right]+1)^2\times K
\label{eq:3}
\end{equation}
where $p$ is the size of the \textit{padding}, $s$ is the size of the \textit{stride}, and $K$ is the number of kernels. Given the above input layer, without any filtering of the non-receptive field values, the number of required microrings would be,
\begin{equation}
N_{microrings} = N_{input}\times K\times N_{kernel}
\label{eq:rings_worse}
\end{equation}
By filtering the non-receptive field values, the total number of rings will drop to:
\begin{equation}
N_{microrings} = K\times N_{kernel}
\label{eq:rings_better}
\end{equation}
One important takeaway from equation \ref{eq:rings_better} is that unlike equation \ref{eq:rings_worse} where the total number of rings scales with product of input size, number of kernels, and the kernel size, here the total number of rings scales linearly with the number of kernels \textbf{$K$} and its size \textbf{$N_{kernel}$}.

For instance, the first convolutional layer of \textit{AlexNet} with an input feature map of shape $224\times 224\times 3$ and 96 kernels of shape $11\times 11\times 3$ will require approximately $5.2$ Billion microrings without filtering non-receptive field values. However, the same number once non-receptive field values are filtered would be $35$ thousand. The latter translates into a saving of more than $150k\times$ in the number microrings. Similarly, the 4th layer of \textit{AlexNet}, which accounts for the most number of kernel weights will require $3456$ microrings. Considering a microring size of $25\mu m \times 25\mu m$ \cite{tait2017neuromorphic}, it takes an area of $2.2mm^2$ to fit all the microrings. Figure \ref{fig:rings_concept} compares the number of microring for different layers of \textit{AlexNet}.
\begin{figure}[ht]
    \centering
    \includegraphics[scale=0.25]{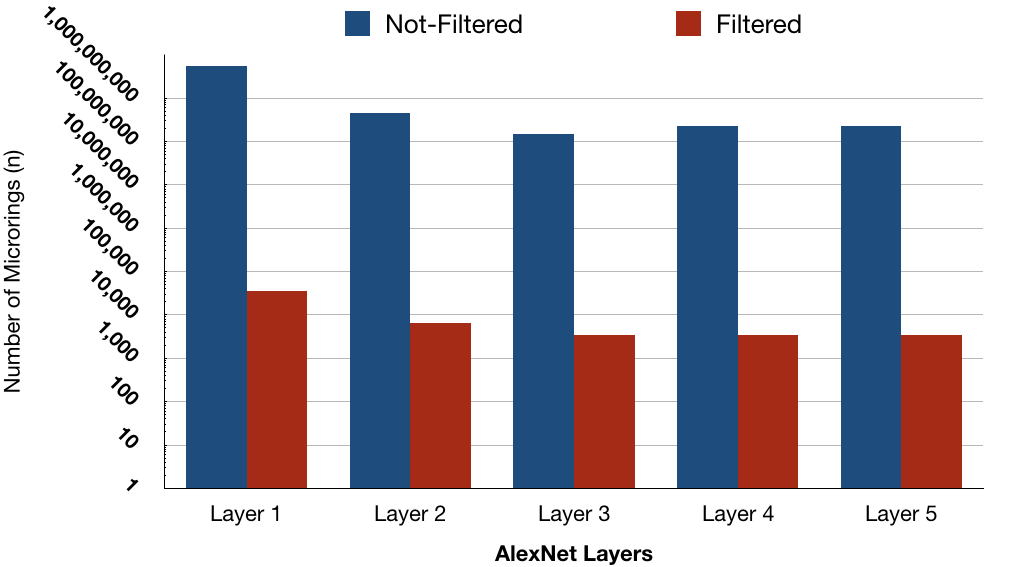}
    \caption{Total number of microrings required in MRR banks for different convolutional layers of \textit{AlexNet} for two cases, namely redundant microrings Filtered and Not-Filtered.} 
    \label{fig:ring_saving}
\end{figure}

\subsection{Execution Time}
Here we derive an analytical estimation of the execution time for the PCNNA. As discussed in section \ref{sec:design_method} The optical convolution core of PCNNA is capable of computing convolutions of multiple kernels in parallel for a single receptive field within a single clock cycle. We name this time window $T_{mac\_i}$, which equals the time to complete all multiply-and-accumulate operations for a series of kernels over the receptive field location $i_{th}$. The number of locations ($N_{locs}$) kernels can have on the input feature maps is found by equation \ref{eq:3} divided by its number of channels,

\begin{equation}
N_{locs} = \frac{N_{output}}{K} = (\left[\frac{(n+2p-m)}{s}\right]+1)^2
\end{equation}

Therefore, without considering the electronic IO limitations, the computation time to perform a full convolution for for an input feature map and $K$ kernels is,
\begin{equation}
T_{conv} = N_{locs}\times \frac{1}{f_{clock}}
\label{eq:conv_time}
\end{equation}

\begin{figure}[ht]
    \centering
    \includegraphics[scale=0.24]{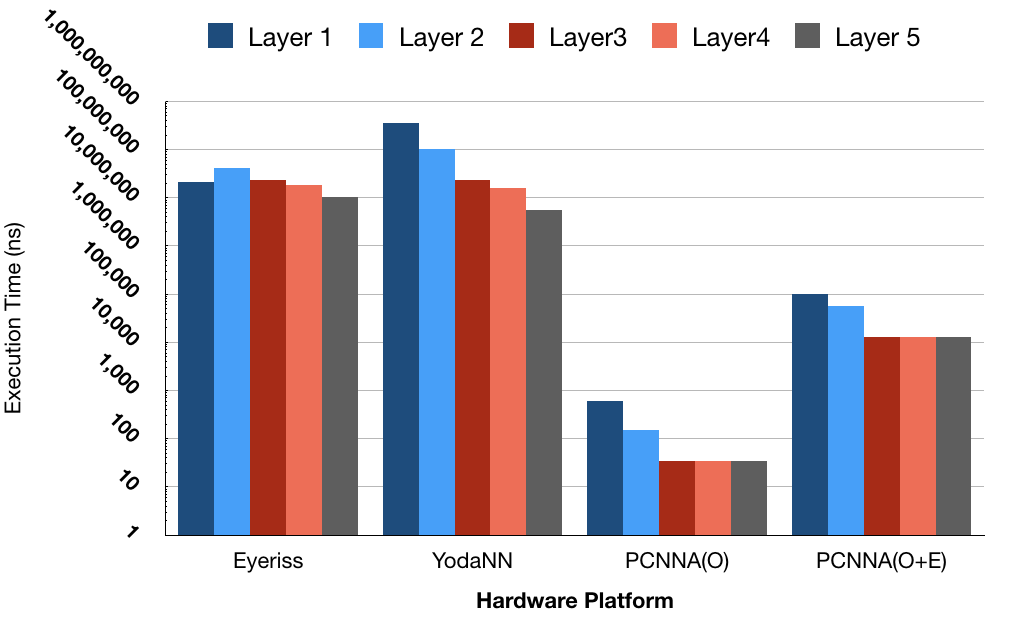}
    \caption{Comparison of execution time for convolution layers of \textit{AlexNet}. PCNNA(O) indicates the Optical core of PCNNA without electronic IO constraints. PCNNA(O+E) indicates full system comprising Optical and Electronic sub-systems.}
    \label{fig:execution_time}
\end{figure}

It should be noted that $T_{conv}$ in equation \ref{eq:conv_time} is independent of the number of kernels. This allows for increasing the number of kernels without sacrificing execution time. The only overhead associated with increased number of kernels in PCNNA is the allocation of more dedicated microrings per kernel. However, the number of microrings increase only linearly with the number of kernels. The total execution times based on a $5GHz$ clock for each layer of \textit{Alexnet} using PCNNA is listed in Figure \ref{fig:execution_time}. This result shows that the PCNNA core has the potential of speedups of up to 5 orders of magnitude in comparison to its cutting-edge electronic counterparts. But, a full system implementation performance of PCNNA is bound by the electronics, both at the front-end and the back-end.
On the front-end, for each iteration of kernels across the input feature map, the corresponding receptive field values are loaded from the off-chip DRAM into the buffer. For each convolution layer, the first location of kernels on the input feature map require to load $N_{kernel}=m\times m\times n_{c}$ value into the buffer. However, as the kernel moves across the input feature map, for subsequent locations of kernels, only a subset of $N_{kernel}$ values equal to $n_c\times s$ must be updated. The stride value $s$ is usually $1$ and in general smaller stride values are preferred over larger ones as they tend to retain more information from at the boundaries of kernels locations on inputs.

Buffered inputs are cached in the SRAM memory \cite{fukuda201413}, which has a $128kb$ capacity that can store $8$ thousand $16$bit values. The access time for the memory is $7ns$ and it has a footprint of $0.443mm^2$. Cached values need to be converted to analog signals using Digital-to-Analog Converters (DAC). In PCNNA DACs operate at a rate of $6GSa/s$ \cite{lin201816b} while each takes up an area of $0.52mm^2$. Our design comprises 1 kernel weight DAC and 10 input DACs. It is worth noting that for every single set of kernel weights for a CNN layer, multiple input values need to be loaded corresponding to different locations of kernels over the input feature map. Here, analog input values from DAC modulate the laser beams with  Mach Zehnder Modulators (MZM), which are usually faster than the $5GHz$ clock.

At the output, calculated convolutions are digitized with a $2.8GSa/s$ Analog-to-Digital Converter (ADC) \cite{stepanovic20132} and stored into the off-chip DRAM through the output buffer. Considering all hardware specifications, the speed bottleneck of PCNNA is the DAC. For every location of kernels over the input feature map a DAC needs to sequentially convert digital inputs to analog values at each stride. This number for largest layer of \textit{AlexNet} with a stride of 1 and 10 ($N_{DAC}$) DACs equals:
\begin{equation}
N_{updated-inputs}=\frac{n_c\times m\times s}{N_{DAC}}=\frac{384\times 3\times 1}{10}\approx 116
\end{equation}
We calculated the execution time using this speed constraint. Figure \ref{fig:execution_time} reports on the estimated execution time of convolution layers of \textit{AlexNet} on PCNNA in comparison with \textit{Eyeriss} \cite{chen2016eyeriss} and \textit{YodaNN} \cite{andri2016yodann}. In Figure \ref{fig:execution_time}, PCNNA(O) indicates only the purely optical core execution times and PCNNA(O+E) represent the full system estimated execution times considering the electronic constraints. Even with electronic IO speed restrictions the full PCNNA system still has the potential of saving up to more than 3 orders of magnitude in execution time. 

\section{Conclusions}
In this paper we presented a proof of concept photonic accelerator for convolutional neural networks. Our proposed accelerator is based the broadcast-and-weight protocol, which takes advantage of microring weight banks to perform multiply and accumulate operations. In PCNNA, we exploited the sparsity of connections between input feature maps and kernels to reduce the number of microrings required to implement modern convolutional neural networks. We showed that the PCNNA optical core has the potential of achieving speedups of up to 5 orders of magnitude. However, electronic IO impose bandwidth limitations to efficiently transfer input data to the optical core. Yet, even when taking these electronic I/O limitations into account, we this optical accelerator shows a 3 orders of magnitude execution time improvement over electronic engines.

\bibliographystyle{IEEEtran}
\bibliography{./references.bib}

\end{document}